\documentclass[preprint,prd,nofootinbib]{revtex4}
\usepackage[dvips]{graphicx}
\usepackage{latexsym}
\newcommand{\nc}{\newcommand}
\nc{\beq}{\begin{equation}}
\nc{\eeq}{\end{equation}}
\nc{\beqa}{\begin{eqnarray}}
\nc{\eeqa}{\end{eqnarray}}
\nc{\bert}{\raise-.55mm\hbox{\large$\Box$}} %D'Alambertian
  \def\lsim{\mathrel{\rlap{\lower4pt\hbox{\hskip1pt$\sim$}}
\raise1pt\hbox{$<$}}}

\def\gsim{\mathrel{\rlap{\lower4pt\hbox{\hskip1pt$\sim$}}
    \raise1pt\hbox{$>$}}}

\def\lsim{\mathrel{\rlap{\lower4pt\hbox{\hskip1pt$\sim$}}
    \raise1pt\hbox{$<$}}}

\long\def\ata#1{\footnote{E-mail: {\tt #1}}}
\long\def\adr#1{$^{\rm #1}$}

\long\def\sl#1{\hbox{\tiny \it #1}}

\newwrite\ffile\global\newcount\figno \global\figno=1

\def\writedef#1{}
\def\figin{\epsfcheck\figin}\def\figins{\epsfcheck\figins}

\def\epsfcheck{\ifx\epsfbox\UnDeFiNeD \message{(NO epsf.tex, FIGURES
WILL BE IGNORED)}
\gdef\figin##1{\vskip2in}\gdef\figins##1{\hskip.5in}% blank space
instead \else\message{(FIGURES WILL BE INCLUDED)}%
\gdef\figin##1{##1}\gdef\figins##1{##1}\fi} \def\figinsert{}
\def\ifig#1#2#3{\xdef#1{fig.~\the\figno} \writedef{#1\leftbracket
fig.\noexpand~\the\figno}%
\figinsert\figin{\centerline{#3}}\medskip\centerline{\vbox{\baselineskip12pt
\advance\hsize by -1truein\center\scriptsize{  Fig.~\the\figno.} #2}}
\bigskip\endinsert\global\advance\figno by1}
\def\endinsert{}
\begin{document}

\preprint{CALT-68-2513} \preprint{OITS-756}

\title{~\\ Anthropic Distribution for Cosmological Constant \\ and
Primordial Density Perturbations \bigskip}

\author{Michael~L.~Graesser\adr{a}\ata{graesser@theory.caltech.edu},
Stephen~D.H.~Hsu\adr{b}\ata{hsu@duende.uoregon.edu},
Alejandro~Jenkins\adr{a}\ata{jenkins@theory.caltech.edu},
Mark~B.~Wise\adr{a}\ata{wise@theory.caltech.edu}}

\affiliation{\adr{a}California Institute of Technology, Pasadena, CA 91125}

\affiliation{\adr{b}Institute of Theoretical Science and Department of
Physics, University of Oregon, Eugene, OR 97403 \vspace{6 mm}}

\begin{abstract} \bigskip

The anthropic principle has been proposed as an explanation for the
observed value of the cosmological constant. Here we revisit this
proposal by allowing for variation between universes in the amplitude
of the scale-invariant primordial cosmological density perturbations.
We derive {\it a priori} probability distributions for this amplitude
from toy inflationary models in which the parameter of the inflaton
potential is smoothly distributed over possible universes.  We find
that for such probability distributions, the likelihood that we live
in a typical, anthropically-allowed universe is generally quite small.

\end{abstract}

\maketitle

\setcounter{footnote}{0} \setcounter{page}{2} \setcounter{section}{0}
\setcounter{subsection}{0} \setcounter{subsubsection}{0}

%%%%%%%%%%%%%%%%%%%%%%%%%%%%%%%%%%%%%%%%%%%

\newpage

The Anthropic Principle has been proposed as a possible solution to
the two cosmological constant problems: why the cosmological constant
$\Lambda$ is orders of magnitude smaller than any theoretical
expectation, and why it is non-zero and comparable today to the energy
density in other forms of matter
\cite{linde&banks,barrowtipler,weinberg1}.  This anthropic argument,
which predates direct cosmological evidence of the dark energy, is the
only theoretical prediction for a small, non-zero $\Lambda$
\cite{weinberg1,weinberg2}.  It is based on the observation that the
existence of life capable of measuring $\Lambda$ requires a universe
with cosmological structures such as galaxies or clusters of stars.  A
universe with too large a cosmological constant either doesn't develop
any structure, since perturbations that could lead to clustering have
not gone non-linear before the universe becomes dominated by
$\Lambda$, or else has a very low probability of exhibiting
structure-forming perturbations, because such perturbations would have
to be so large that they would lie in the far tail-end of the cosmic
variance. The existence of the string theory landscape, in which
causally disconnected regions can have different cosmological and
particle physics properties, adds support to the notion of an
anthropic rule for selecting a vacuum.

How well does this principle explain the observed value of $\Lambda$
in our universe? Careful analysis by \cite{weinberg2} finds that 5$\%$
to 12$\%$ of universes would have a cosmological constant smaller than
our own. In everyday experience we encounter events at this level of
confidence,\footnote{For instance, drawing two pairs in a poker hand.}
so as an explanation this is not unreasonable.

If the value of $\Lambda$ is not fixed {\it a priori}, then one might
expect other fundamental parameters to vary between universes as well.
This is the case if one sums over wormhole configurations in the path
integral for quantum gravity \cite{coleman}, as well as in the string
theory landscape \cite{stringlandscape,douglas,others,banksdine}. In
\cite{banksdine} it was emphasized that all the parameters of the low
energy theory would vary over the space of vacua (``the landscape'').
Douglas \cite{douglas} has initiated a program to quantify the
statistical properties of these vacua, with additional contributions
by others \cite{others}.

In \cite{aguirre}, Aguirre stressed that life might be possible in
universes for which some of the cosmological parameters are orders of
magnitude different from those of our own universe.  The point is that
large changes in one parameter can be compensated by changes in
another in such a way that life remains possible.  Anthropic
predictions for a {\it particular} parameter value will therefore be
weakened if other parameters are allowed to vary between
universes. One cosmological parameter that may significantly affect
the anthropic argument is $Q$, the standard deviation of the amplitude
of primordial cosmological density perturbations.  Rees \cite{rees}
and Tegmark and Rees \cite{tegmark} have pointed out that if the
anthropic argument is applied to universes where $Q$ is not fixed but
randomly distributed, then our own universe becomes less likely
because universes with both $\Lambda$ and $Q$ larger than our own are
anthropically allowed. The purpose of this letter is to quantify this
expectation within a broad class of inflationary models.  Restrictions
on the {\it a prori} probability distribution for $Q$ necessary for
obtaining a successful anthropic prediction for $\Lambda$, were
considered in \cite{garrigaliviovilenkin,garrigavilenkin2}.

In our analysis we let both $\Lambda$ and $Q$ vary between universes
and then quantify the anthropic likelihood of a positive cosmological
constant less than or equal to that observed in our own universe.  We
offer a class of toy inflationary models that allow us to restrict the
{\it a priori} probability distribution for $Q$, making only modest
assumptions about the behavior of the {\it a priori} distribution for
the parameter of the inflaton potential in the anthropically-allowed
range. Cosmological and particle physics parameters other than
$\Lambda$ and $Q$ are held fixed as initial conditions at
recombination.  We provisionally adopt Tegmark and Rees's anthropic
bound on $Q$: a factor of 10 above and below the value measured in our
universe.  Even though this interval is small, we find that the
likelihood that our universe has a typical cosmological constant is
drastically reduced. The likelihood tends to decrease further if
larger intervals are considered.

Weinberg determined in \cite{weinberg1} that, in order for an
overdense region to go nonlinear before the energy density of the
universe becomes dominated by $\Lambda$, the value of the overdensity
$\delta \equiv \delta \rho /\rho$ must satisfy \beq \delta >
\left({729 \Lambda \over 500 \bar{\rho}}\right)^{1/3}
~. \label{deltamin} \eeq In a matter-dominated universe this relation
has no explicit time dependence. Here $\bar{\rho}$ is the energy
density in non-relativistic matter.  Perturbations not satisfying the
bound cease to grow once the universe becomes dominated by the
cosmological constant.  For a fixed amplitude of perturbations, this
observation provides an upper bound on the cosmological constant
compatible with the formation of structure. Throughout our analysis we
assume that at recombination $\Lambda \ll \bar{\rho}$.

To quantify whether our universe is a typical, anthropically-allowed
universe, additional assumptions about the distribution of
cosmological parameters and the spectrum of density perturbations
across the ensemble of universes are needed.

A given slow-roll inflationary model with reheating leads to a
Friedman-Roberston-Walker universe with a (late-time) cosmological
constant $\Lambda$ and a spectrum of perturbations that is
approximately scale-invariant and Gaussian with a variance \beq Q^2
\equiv \langle \tilde{\delta}^2 \rangle_{\sl{HC}} ~. \eeq The
expectation value is computed using the ground state in the
inflationary era and perturbations are evaluated at
horizon-crossing. The variance is fixed by the parameters of the
inflationary model together with some initial conditions.  Typically,
for single-field $\phi$ slow-roll inflationary models, \beq Q^2 \sim
\left. { H^4 \over \dot{\phi}^2 }\right\vert _{\sl{HC}}~.  \eeq This
leads to spatially separated over- or underdense regions with an
amplitude $\delta$ that for a scale-invariant spectrum are distributed
(at recombination) according to \beq \label{N} {\cal N}(\sigma,\delta)
= \sqrt{2 \over \pi} {1 \over \sigma} e^{- \delta^2/2 \sigma^2 } ~.
\eeq (The linear relation between $Q$ and the filtered $\sigma$ in Eq.
(\ref{N}) is discussed below.)

By Bayes's theorem, the probability for an anthropically-allowed
universe (i.e. the probability that the cosmological parameters should
take certain values, given that life has evolved to measure them) is
proportional to the product of the {\it a priori} probability
distribution $P$ for the cosmological parameters, times the
probability that intelligent life would evolve given that choice of
parameter values.  Following \cite{weinberg2}, we estimate that second
factor as being proportional to the mean fraction ${\cal
F}(\sigma,\Lambda)$ of matter that collapses into galaxies. The latter
is obtained in a universe with cosmological parameters $\Lambda$ and
$\sigma$ by spatially averaging over all over- or underdense regions,
so that \cite{weinberg2} \beq {\cal F}(\sigma, \Lambda)=
\int^{\infty}_{\delta_{\sl{min}}} d \delta \, N(\sigma, \delta) {\cal
F}(\delta,\Lambda) ~.  \eeq The lower limit of integration is provided
by the anthropic bound of Eq.  (\ref{deltamin}), which gives
$\delta_{\sl{min}} \equiv (729 \Lambda /500 \bar{\rho})^{1/3} $.  The
anthropic probability distribution is

\beq \label{calP} {\cal P}(\sigma,\Lambda) = P(\Lambda,\sigma) {\cal
F}(\sigma, \Lambda)d\Lambda \,d\sigma ~.  \eeq

Computing the mean fraction of matter collapsed into structures
requires a model for the growth and collapse of inhomogeneities.  The
Gunn-Gott model \cite{peebles,gunngott} describes the growth and
collapse of an overdense spherical region surrounded by a compensating
underdense shell. The weighting function ${\cal F}(\delta,\Lambda)$
gives the fraction of mass in the inhomogeneous region of density
contrast $\delta$ that eventually collapses (and then forms
galaxies). To a good approximation it is given by \cite{weinberg2}
\beq {\cal F}(\delta,\Lambda) =\delta {1 \over \delta +
\delta_{\sl{min}}}~. \eeq Additional model-dependence occurs in the
introduction of the parameter $s$ given by the ratio of the volume of
overdense sphere to the volume of the under-dense shell surrounding
the sphere. We will set $s=1$ throughout.

Since the anthropically allowed values for $\Lambda$ are so much
smaller than any other mass scale in particle physics, and since we
assume that $\Lambda=0$ is not a special point in the landscape, we
follow \cite{vilenkin,weinberg2} in using the approximation
$P(\Lambda) \simeq P(\Lambda = 0)$ for $\Lambda$ within the
anthropically allowed window.\footnote{Garriga and Vilenkin point to
examples of quintessence models in which the approximation
\mbox{$P(\Lambda)\simeq P(\Lambda=0)$} in the anthropically-allowed
range is not valid \cite{garrigavilenkin1}.}  The requirement that the
universe not recollapse before intelligent life has had time to evolve
anthropically rules out large negative $\Lambda$
\cite{barrowtipler,weinberg3}. We will assume that the anthropic
cutoff for negative $\Lambda$ is close enough to $\Lambda = 0$ that
all $\Lambda < 0$ may be ignored in our calculations.

As an example of a concrete model for the variation in $Q$ between
universes, we consider inflaton potentials of the form (see, for
example, \cite{kolbturner}) \beq \label{inflatonV} V= \Lambda +
\lambda \phi^{2p}~, \eeq where $p$ is a positive
integer.\footnote{Recent analysis of astronomical data disfavors the
$\lambda \phi^4$ inflationary model \cite{SDSS}, but for generality we
will consider an arbitrary $p$ in Eq.  (\ref{inflatonV}).} We assume
there are additional couplings that provide an efficient reheating
mechanism, but are unimportant for the evolution of $\phi$ during the
inflationary epoch.  The standard deviation of the amplitude of
perturbations gives \beq Q = A \sqrt{\lambda} \,
\frac{\phi^{p+1}_{\sl{HC}}}{M^3_{\sl{Pl}}}~, \label{Qlambda} \eeq
where $A$ is a constant, and $\phi_{\sl{HC}}$ is the value of the
field when the mode of wave number $k$ leaves the horizon.  This
$\phi_{\sl{HC}}$ has logarithmic dependence on $\lambda$ and $k$,
which we neglect.  Randomness in the initial value for $\phi$ affects
only those modes that are (exponentially) well outside our
horizon. Throughout this letter, we will set the spectral index to one
and ignore its running. Equation (\ref{Qlambda}) then gives $\lambda
\propto Q^2$.

Next, suppose that the fundamental parameters of the Lagrangrian are
not fixed, but vary between universes, as might be expected if one
sums over wormhole configurations in the path integral for quantum
gravity \cite{coleman} or in the string theory landscape
\cite{stringlandscape,douglas,banksdine,others}.  To obtain the
correct normalization for the density perturbations observed in our
universe, the self-coupling must be extremely small. As the standard
deviation $Q$ will be allowed to vary by an order of magnitude around
$10^{-5}$, for this model the self-coupling in alternate universes
will be very small as well.

We may then perform an expansion about $\lambda = 0$ for the {\it a
priori} probability distribution of $\lambda$.  The smallness of
$\lambda$ suggests that we may keep only the leading term in that
expansion.  If the {\it a priori} probability distribution extends to
negative values of $\lambda$ (which are anthropically excluded due to
the instability of the resulting action for $\phi$), we expect it to
be smooth near $\lambda=0$, and the leading term in the power series
expansion to be zeroth order in $\lambda$ (i.e., a constant).
Therefore we expect a flat {\it a priori} probability distribution for
$\lambda$. The {\it a priori} probability distribution for $Q $ is
then \beq P(Q) \propto {d \lambda \over d Q} \sim Q~, \eeq where the
normalization constant is determined by the range of integration in
$Q$. Note that this distribution favors large $Q$.  On the other hand,
if the {\it a priori} probability distribution for the coupling
$\lambda$ only has support for $\lambda > 0$ then $\lambda=0$ is a
special point and we cannot argue that $P(Q) \propto Q$.  However,
since the anthropically-allowed values of $\lambda$ are very small,
the {\it a priori} distribution for $\lambda$ should be dominated, in
the anthropically-allowed window, by a leading term such as
$P(\lambda) \sim \lambda^q$.  Normalizability requires $q>-1$.  Using
$\lambda \propto Q^2$, this gives $P(Q)\sim Q^{2q+1}$.

Before proceeding, it is convenient to transform to the new variables:
\beq y \equiv {\Lambda \over \rho_\ast} ~~;~~ \hat{\sigma} \equiv
\sigma \left( { \bar{\rho} \over \rho_{\ast}} \right)^{1 \over 3}
~. \eeq Here $\bar{\rho}$ is the energy-density in non-relativistic
matter at recombination, which we take to be fixed in all universes,
and $\rho_\ast$ is the value for the present-day energy density of
non-relativistic matter in our own universe. For a matter-dominated
universe $\hat{\sigma}$ is time-independent, whereas $y$ is constant
for any era.  Here and throughout this letter, a subscript $\ast$
denotes the value that is observed in our universe for the
corresponding quantity.  The only quantities whose variation from
universe to universe we will consider are $y$ and $\hat{\sigma}$.

In terms of these variables and following \cite{weinberg2}, the
probability distribution of Eq.  (\ref{calP}) is found to be \beq
{\cal P} = N d\hat{\sigma} dy \, P(\hat{\sigma}) \int
^{\infty}_{\beta} dx {e^{-x} \over \beta^{1/2} + x^{1/2}}
\label{pdffinal}~, \eeq where \beq \beta \equiv {1 \over 2
\hat{\sigma}^2} \left({729 y \over 500}\right)^{2 /3}~, \eeq and $N$
is the normalization constant.

Notice that, since $x \geq \beta$, large $\beta$ implies that ${\cal
P} \sim e^{-\beta} \ll 1$.  For a fixed $\hat{\sigma}$, large $y$
implies large $\beta$.  Thus, for fixed $\hat{\sigma}$, large
cosmological constants are anthropically disfavored.  But if
$\hat{\sigma}$ is allowed to increase, then $\beta \sim {\cal O}(1)$
may be maintained at larger $y$.  Garriga and Vilenkin have pointed
out that the distribution in Eq.  (\ref{pdffinal}) may be rewritten
using the change of variables $(\hat \sigma, y) \mapsto (\hat \sigma,
\beta)$ \cite{garrigavilenkin2}.  The Jacobian for that transformation
is a function only of $\hat{\sigma}$.  Equation (\ref{pdffinal}) then
factorizes into two parts: one depending only on $\hat{\sigma}$, the
other only on $\beta$.  Integration over $\hat{\sigma}$ produces an
overall multiplicative factor that cancels out after normalization, so
that any choice of $P(\hat{\sigma})$ will give the same distribution
for the dimensionless parameter $\beta$.  In that sense, even in a
scenario where $\hat{\sigma}$ is randomly distributed, the computation
in \cite{weinberg2} may be seen as an anthropic prediction for
$\beta$.\footnote{We thank Garriga and Vilenkin for explaining this
point to us.} The measured value of $\beta$ is, indeed, typical of
anthropically-allowed universes, but an anthropic explanation for
$\beta$ alone does not address the problem of why both $\Lambda$ and
$Q$ should be so small in our universe.

Implementing the anthropic principle requires making an assumption
about the minimum mass of ``stuff'', collapsed into stars, galaxies,
or clusters of galaxies, that is needed for the formation of life.  It
is more convenient to express the minimum mass $M_{\sl{min}}$ in terms
of a comoving scale $R$: $M_{\sl{min}}= 4 \pi \bar{\rho} a_{eq}^3
R^3/3 $ (by convention $a=1$ today, so $R$ is a physical scale). We do
not know the precise value of $R$.  A better understanding of biology
would in principle determine its value, which should only depend on
chemistry, the fraction of matter in the form of baryons, and Newton's
constant.  In our analysis these are all fixed initial conditions at
recombination. In particular, we would not expect $M_{\sl{min}}$ to
depend on $\Lambda$ or $Q$.\footnote{Note, however, that requiring
life to last for billions of years (long enough for it to develop
intelligence and the ability to do astronomy) might place bounds on
$Q$.  See \cite{tegmark}.} Therefore, even though the relation between
$M_{\sl{min}}$ and $R$ depends on present-day cosmological parameters,
the value of this threshold will be constant between universes because
it depends only on parameters that we are treating as fixed initial
conditions.  Thus, in computing the probability distribution over
universes, we will fix $R$.  Since we don't know what is the correct
anthropic value for $R$, we will present our results for both $R$=1
and 2 Mpc.  ($R$ on the order of a few Mpc corresponds to requiring
that structures as large as our galaxy be necessary for life.)

We then proceed to filter out perturbations with wavelength smaller
than $R$, leading to a variance $\sigma^2$ that depends on the
filtering scale.  Expressed in terms of the power spectrum evaluated
at recombination, \beq \sigma^2 = {1 \over 2 \pi^2} \int^{\infty}_0 dk
\, k^2 P(k) W^2(kR) \label{sigmaR} \eeq where $W$ is the filter
function, which we take to be a Gaussian $W(x)=e^{-x^2/2}$.  $P(k)$ is
the power spectrum, which we assume to be scale-invariant.  (For
$P(k)$ we use Eq. (39) of \cite{weinberg2}, setting $n=1$).

Evaluating (\ref{sigmaR}) at recombination gives, for our universe,
\beq \hat{\sigma}_\ast = C_\ast Q_{\ast} ~. \eeq The number $C_\ast$
contains the growth factor and transfer function evaluated from
horizon crossing to recombination and only depends on physics from
that era.  We assume $\Lambda$ is small enough so that at
recombination it can be ignored and thus we take the variation in
$\hat{\sigma}$ between universes to come solely from its explicit
dependence on $Q$.

We may then use observations of $Q_\ast$ and $\sigma_\ast$ to
determine $\hat{\sigma} = C_\ast Q$, valid for all universes.  We use
the explicit expression for $C_\ast$ that is obtained from
Eqs. (39)-(43) and (48)-(51) in Ref. \cite{weinberg2}.  This takes as
inputs the Hubble parameter $H_0\equiv 100 h_\ast $km/s, the energy
density in non-relativistic matter $\Omega_\ast$, the cosmological
constant $\lambda_\ast=1- \Omega_\ast$, the baryon fraction $\Omega_b
=0.023 h_\ast^{-2}$, the smoothing scale $R$, and the COBE normalized
amplitude of fluctuations at horizon crossing, $Q_\ast=
1.94\times10^{-5} \Omega_\ast^{-.785 - 0.05\ast\ln\Omega_\ast}$.

As we have argued, the dependence of $C_\ast$ on the cosmological
constant is not relevant for our purposes.  For our calculations we
use $\Omega_\ast=0.134 h_\ast^{-2}$, and $h_\ast=0.73$ (consistent
with their observed best-fit values \cite{pdg}).  The smoothing scale
$R$ will be taken to be either 1 Mpc or 2 Mpc, and the corresponding
values for $C_\ast$ are $5.2 \cdot 10^4$ and $3.8 \cdot 10^4$.

The values chosen for the range of $Q$ are motivated by the discussion
in \cite{tegmark} about anthropic limits on the amplitude of the
primordial density perturbations.  The authors of \cite{tegmark} argue
that $Q$ between $10^{-3}$ and $10^{-1}$ leads to the formation of
numerous supermassive blackholes which might obstruct the emergence of
life.\footnote{They also note that for $Q> 10^{-4}$ formation of life
is possible, but planetary disruptions caused by flybys may make it
unlikely for planetary life to last billions of years.} They then
claim that universes with $Q$ less than $10^{-6}$ are less likely to
form stars, or if star clusters do form, that they would not be bound
strongly enough to retain supernova remnants.  Since there is
considerable uncertainty in these limits, we carry out calculations
using both the range indicated by \cite{tegmark} as well as a range
that is somewhat broader.\footnote{Notice that we are using the ranges
indicated in \cite{tegmark} as absolute anthropic cutoffs.  Arguments
like those made in \cite{tegmark} introduce some correction to the
approximation made in \cite{weinberg2} that the probability of life is
proportional to the amount of matter that collapses into compact
structures.  Since we are largely ignorant of what the form of this
correction is, we have approximated it as a simple window function.}

Previous work on applying the anthropic principle to variable
$\Lambda$ and $Q$ has assumed {\it a priori} distributions $P(Q)$ that
fall off as $1/Q^k$ for large $Q$, with $k\geq 3$
\cite{garrigaliviovilenkin,garrigavilenkin2}.  Such distributions were
chosen in order to keep the anthropic probability ${\cal P}(y,Q)$
normalizable, and they usually yield anthropic predictions for the
cosmological constant similar to those that were obtained in
\cite{weinberg2} by fixing $Q$ to its observed value, because they
naturally favor a $Q$ as small as its observed value in our universe.
For instance, for $P(Q) \propto 1/Q^3$ in the range $Q_\ast/10 < Q <
10 Q_\ast$, $P(y < y_\ast) = 5\%$ for $R=1$ Mpc, while $P(y < y_\ast)
= 7\%$ for $R=2$ Mpc.)

However, if we accept the argument of Tegmark and Rees in
\cite{tegmark} that there are natural anthropic cutoffs on $Q$, it
follows that the behavior of $P(Q)$ at large $Q$ is irrelevant to the
normalizability of ${\cal P}(y,Q)$.  Furthermore, $P(Q)\sim 1/Q^k$ in
the neighborhood of $Q=0$ for $k\geq1$ leads to an unnormalizable
distribution, since the integral $\int P(Q) dQ$ blows up.  In what
follows we shall consider two {\it a priori} distributions: $P(Q)
\propto Q$, and $P(Q)\propto 1/Q^{0.9}$ inside the anthropic window,
motivated by the inflationary models we have discussed.

The results are summarized in Table \ref{y-table}, where $P(y <
y_\ast)$ is the anthropic probability that the value $y$ be no greater
than what is observed in our own universe, $\langle y \rangle$ is the
anthropically-weighed mean value of $y$, and $y_{\sl{5\%}}$ is the
value of $y$ such that the anthropic probability of obtaining a value
no greater than that is 5\%.

\begin{table}[t] \fontsize{10}{12} \selectfont \centerline{
\begin{tabular}{| c || c | c | c | c | c | c | c | c |} \hline
 &\multicolumn{4}{c|}{$P(Q)\propto 1/Q^{0.9}$ in the range}
 &\multicolumn{4}{|c|}{$P(Q)\propto Q$ in the range} \\ \cline{2-9}
 &\multicolumn{2}{c|}{$Q_\ast/10 < Q < 10Q_\ast$} &
 \multicolumn{2}{|c|}{$Q_\ast/15 < Q < 15Q_\ast$}
 &\multicolumn{2}{|c|}{$Q_\ast/10 < Q < 10Q_\ast$} &
 \multicolumn{2}{|c|}{$Q_\ast/15 < Q < 15Q_\ast$}\\ \cline{2-9} &
 {$R=1$ Mpc} & {$R=2$ Mpc} & {$R=1$ Mpc} & {$R=2$ Mpc} & {$R=1$ Mpc} &
 {$R=2$ Mpc} & {$R=1$ Mpc} & {$R=2$ Mpc} \\ \hline \hline $P(y <
 y_\ast)$ & 1 $\cdot 10^{-3}$ & 3 $\cdot 10^{-3}$ & 4$\cdot 10^{-4}$ &
 1 $\cdot 10^{-3}$& 5 $\cdot 10^{-4}$ & 1 $\cdot 10^{-3}$ & 1 $\cdot
 10^{-4}$ & 4 $\cdot 10^{-4}$ \\ \hline $\langle y \rangle/y_\ast$ & 1
 $\cdot 10^4 $ & 4 $\cdot 10^3 $ & 4 $\cdot 10^4 $ & 1 $\cdot 10^4$ &
 1 $\cdot 10^4 $ & 5 $\cdot 10^3 $ & 4 $\cdot 10^4 $ & 2 $\cdot 10^4 $
 \\ \hline $y_{\sl{5\%}}/y_\ast$ & 9 $\cdot 10$ & 4 $\cdot 10$ &
 3$\cdot 10^2$ &1 $ \cdot 10^2$ & 2 $ \cdot 10^2$ & 7$\cdot 10$ &
 6$\cdot 10^2$ & 2 $\cdot 10^2$ \\ \hline
\end{tabular}}
\caption{Anthropically Determined Properties of the Cosmological
  Constant}
\label{y-table}
\end{table}

\begin{table}[t] \fontsize{10}{12} \selectfont \centerline{ 
\begin{tabular}{| c || c | c | c | c | c | c | c | c |} \hline
 &\multicolumn{4}{c|}{$P(Q)\propto 1/Q^{0.9}$ in the range}
 &\multicolumn{4}{|c|}{$P(Q)\propto Q$ in the range} \\ \cline{2-9}
 &\multicolumn{2}{c|}{$Q_\ast/10 < Q < 10Q_\ast$} &
 \multicolumn{2}{|c|}{$Q_\ast/15 < Q < 15Q_\ast$}
 &\multicolumn{2}{|c|}{$Q_\ast/10 < Q < 10Q_\ast$} &
 \multicolumn{2}{|c|}{$Q_\ast/15 < Q < 15Q_\ast$}\\ \cline{2-9} &
 {$R=1$ Mpc} & {$R=2$ Mpc} & {$R=1$ Mpc} & {$R=2$ Mpc} & {$R=1$ Mpc} &
 {$R=2$ Mpc} & {$R=1$ Mpc} & {$R=2$ Mpc} \\ \hline \hline $P(Q <
 Q_\ast)$ & 8 $\cdot 10^{-4}$ & 8 $\cdot 10^{-4}$ & 2 $\cdot 10^{-4}$
 & 2 $\cdot 10^{-4}$ & 1 $\cdot 10^{-5}$ & 1 $\cdot 10^{-5}$ & 1
 $\cdot 10^{-6}$ & 1 $\cdot 10^{-6}$ \\ \hline $\langle Q
 \rangle/Q_\ast$ & 8 & 8 & 11 & 11 & 8 & 8 & 13 & 13 \\ \hline
 $Q_{\sl{5\%}}/Q_\ast$ & 4 & 4 & 6 & 6 & 5 & 5 & 8 & 8 \\ \hline
\end{tabular}} 
\caption{Anthropically Determined Properties of the Amplitude for
  Density Pertubations}
\label{Q-table}
\end{table}

By comparison, for this choice of cosmological parameters, the authors
of \cite{weinberg2} find that, for $Q$ fixed (or measured), the
probability of a universe having a cosmological constant no greater
than our own is much higher: $P(y<0.7/0.3)=.05$ and $0.1$, for $R=1$
Mpc and $R=2$ Mpc respectively.\footnote{These numbers are taken from
Table 1 in the published version of Ref. \cite{weinberg2}.}

One can also ask what is the probability of observing a value for $Q$
in the range $Q_\ast/10 < Q < Q_\ast$, after averaging over all
possible cosmological constants.  Table \ref{Q-table} summarizes the
resulting distribution in $Q$.

In summary, inflation and a landscape of anthropically determined
coupling constants provides (in some scenarios) a conceptually clean
framework for variation between universes in the magnitude of
$Q$. Since increasing $Q$ allows the probability of structure to
remain non-negligible for $\Lambda$ considerably larger than in our
own universe, anthropic solutions to the cosmological constant problem
are weakened by allowing $Q$ as well as $\Lambda$ to vary from one
universe to another.\\

\noindent The authors thank J.~Garriga and A.~Vilenkin for valuable
comments.  This work was supported in part under DOE grants
DE-FG06-85ER40224 and DE-FG03-92ER40701.

%%%%%%%%%%%%%%%%%%%%%%%%%%%%%%
%  Bibliography
%%%%%%%%%%%%%%%%%%%%%%%%%%%%%%


\baselineskip=1.6pt
\begin{thebibliography}{99}

\bibitem{linde&banks}A.D.~Linde, in {\it The Very Early Universe},
eds. G.W.~Gibbons et al. (Cambridge University Press, Cambridge,
1983); T.~Banks, Nucl. Phys. {\bf B249}, 332 (1985).

\bibitem{barrowtipler} J.D.~Barrow and F.J.~Tipler, {\it The Anthropic
Cosmological Principle} (Oxford University Press, Oxford, 1986).

\bibitem{weinberg1} S.~Weinberg, Phys. Rev. Lett {\bf 59}, 2067 (1987).

\bibitem{weinberg2} H.~Martel, P.~Shapiro, and S.~Weinberg,
Astrophys. J.  {\bf 492}, 29 (1998) [astro-ph/9701099].

\bibitem{coleman} S.~Coleman, Nucl. Phys. {\bf B307}, 867 (1988);
S.~Giddings and A.~Strominger, Nucl. Phys. {\bf B307}, 854 (1988).

\bibitem{stringlandscape} S.~Kachru, R.~Kallosh, A.~Linde, and
S.P.~Trivedi, Phys. Rev. D {\bf 68}, 0046005 (2003) [hep-th/0301240]; L.
Susskind, hep-th/0302219.

\bibitem{douglas} M.~Douglas, JHEP {\bf 0305}, 046 (2003)
[hep-th/0303194]; hep-th/0401004.

\bibitem{others} A.~Giryavets, S.~Kachru, P.K.~Tripathy, and
S.P.~Trivdedi, JHEP {\bf 0404}, 003, (2003) [hep-th/0312104];
A.~Giryavets, S.~Kachru, and P.K.~Tripathy, hep-th/0404243;
L.~Susskind, hep-th/0405189; M.~Dine, E.~Gorbatov, and S.~Thomas,
hep-th/0407043.

\bibitem{banksdine} T.~Banks, M.~Dine, and E.~Gorbatov, hep-th/0309170.

\bibitem{aguirre} A.~Aguirre, Phys. Rev. D {\bf 64} 083508 (2001)
[astro-ph/0106143].

\bibitem{rees} M.J.~Rees, Complexity {\bf 3} 17 (1997); in {\it Fred
Hoyle's Universe}, eds. C.~Wickramasinghe et al.  (Kluwer, Dordrecht,
2003) [astro-ph/0401424].

\bibitem{tegmark} M.~Tegmark and M.J.~Rees, Astrophys. J. {\bf 499},
  526 (1998) [astro-ph/9709058].

\bibitem{garrigaliviovilenkin}J.~Garriga, M.~Livio, and A.~Vilenkin, Phys.
Rev. D {\bf 61}, 023503 (2000) [astro-ph/9906210].

\bibitem{garrigavilenkin2}J.~Garriga and A.~Vilenkin, Phys. Rev. D {\bf
67}, 043503 (2003) [astro-ph/0210358].

\bibitem{peebles} P.J.E.~Peebles, Astrophys. J. {\bf 147}, 859 (1967).

\bibitem{gunngott} J.E.~Gunn and J.R.~Gott, Astrophys. J. {\bf 176}, 1 (1972).

\bibitem{vilenkin}A.~Vilenkin, gr-qc/9512031.

\bibitem{garrigavilenkin1} J.~Garriga and A.~Vilenkin, Phys. Rev. D {\bf
61} 083502 (2000) [astro-ph/9908115].

\bibitem{weinberg3} S.~Weinberg, in {\it Critical Dialogues in
Cosmology}, ed. N.~Turok (World Scientific, Singapore, 1996)
[astro-ph/9610044].

\bibitem{kolbturner} E.W.~Kolb and M.S.~Turner, {\it The Early
Universe}, (Addison-Wesley, Reading MA, 1990).

\bibitem{SDSS}U.~Seljak et al., astro-ph/0407372.

\bibitem{pdg} S.~Eidelman et al., Phys. Lett. B {\bf 592}, 1 (2004).

\end{thebibliography}
\end{document}